\def\papertitle{Band-Count Dense Modal Estimation with Fixed-Frequency Differentiable Resonator Refinement}
\def\paperauthorA{Minhui Lu}
\def\paperauthorB{Joshua D. Reiss}

\documentclass[twoside,a4paper]{article}
\usepackage{etoolbox}
\usepackage[taskB]{dafx26challenge}

\usepackage{amsmath,amssymb,amsfonts}
\usepackage{booktabs}
\usepackage{siunitx}
\usepackage[T1]{fontenc}
\usepackage[utf8]{inputenc}
\usepackage[english]{babel}
\usepackage{caption}
\AtBeginDocument{\urlstyle{same}}

\input glyphtounicode
\ninept
\newcounter{numauth}
\newcounter{listcnt}
\newcommand\authcnt[1]{\ifdefined#1 \stepcounter{numauth} \fi}
\newcommand\addauth[1]{\ifdefined#1 \stepcounter{listcnt}\ifnum \value{listcnt}<\value{numauth}\appto\authorslist{, #1}\else\appto\authorslist{~and~#1}\fi\fi}
\authcnt{\paperauthorB}
\authcnt{\paperauthorC}
\authcnt{\paperauthorD}
\authcnt{\paperauthorE}
\authcnt{\paperauthorF}
\authcnt{\paperauthorG}
\authcnt{\paperauthorH}
\authcnt{\paperauthorI}
\authcnt{\paperauthorJ}
\def\authorslist{\paperauthorA}
\addauth{\paperauthorB}
\addauth{\paperauthorC}
\addauth{\paperauthorD}
\addauth{\paperauthorE}
\addauth{\paperauthorF}
\addauth{\paperauthorG}
\addauth{\paperauthorH}
\addauth{\paperauthorI}
\addauth{\paperauthorJ}

\usepackage{times}
\newif\ifpdf
\ifx\pdfoutput\relax
\else
   \ifcase\pdfoutput
      \pdffalse
   \else
      \pdftrue
   \fi
\fi

\ifpdf
  \usepackage[pdftex,
    pdftitle={\papertitle},
    pdfauthor={\authorslist},
    pdfsubject={Proceedings of the 29th International Conference on Digital Audio Effects (DAFx26)},
    colorlinks=false,
    bookmarksnumbered,
    pdfstartview=XYZ
  ]{hyperref}
  \usepackage[pdftex]{graphicx}
\else
  \usepackage[dvips]{epsfig,graphicx}
  \usepackage[dvips,
    pdftitle={\papertitle},
    pdfauthor={\authorslist},
    pdfsubject={Proceedings of the 29th International Conference on Digital Audio Effects (DAFx26)},
    colorlinks=false,
    bookmarksnumbered,
    pdfstartview=XYZ
  ]{hyperref}
\fi

\title{\papertitle}
\affiliation
{\paperauthorA~and~\paperauthorB}
{Centre for Digital Music \\ Queen Mary University of London \\ London, United Kingdom\\
{\tt \href{mailto:minhui.lu@qmul.ac.uk}{minhui.lu@qmul.ac.uk},
\href{mailto:joshua.reiss@qmul.ac.uk}{joshua.reiss@qmul.ac.uk}}}

\begin{document}
\ifpdf
  \DeclareGraphicsExtensions{.png,.jpg,.pdf}
\else
  \DeclareGraphicsExtensions{.eps}
\fi
\maketitle

\begin{abstract}
Task B of the 1st DAFx Parameter Estimation Challenge requires estimating the
frequencies, decay rates, gains, and number of modes in a dense plate-reverb
impulse response. Weak and overlapping modes make sparse peak detection prone to
severe undercounting. We train an ExtraTrees regressor on simulator-generated
data to predict mode counts in four frequency bands. These counts define dense
frequency grids, after which a differentiable all-pole resonator model refines
decay and gain while keeping frequency fixed. On two separate synthetic
validation sets, the system reduces a local challenge-style error by about 66\%
relative to the official default peak-picking baseline. The improvement is
mainly associated with lower mode-count mismatch, while decay and gain remain
the largest error sources. These findings support separating modal-density
estimation from continuous parameter fitting.
\end{abstract}

\section{Introduction}

Modal representations are widely used in physical audio modeling and
artificial reverberation because they express a resonant system as a sum of
damped components. For a plate, these components are specified by modal
frequencies, decay rates, and gains, or equivalently by a bank of resonant
filters~\cite{smith2010physical,bilbao2009numerical}. The representation
supports interpretable analysis and efficient resynthesis, but recovering it
from a single dense impulse response is difficult. Modes can be closely spaced,
weak resonances can be masked by stronger ones, and the mode count is unknown.

Task B of the 1st DAFx Parameter Estimation Challenge provides a benchmark for
this problem. The two challenge tasks address different levels of the same
generative process. Task A estimates a compact physical and observation
description of the plate system, whereas Task B directly estimates the much
larger modal representation whose components form the response. Each target
response is generated by a public modal plate-reverb simulator available in the
official challenge repository:\footnote{\url{https://github.com/LOGUNIVPM/1st-DAFx-Challenge}.}
The required output is a modal set
\begin{equation}
\mathcal{M}=\{(f_m,\sigma_m,g_m)\}_{m=1}^{M},
\end{equation}
where $f_m$ is modal frequency, $\sigma_m$ is the decay rate, $g_m$ is the
modal gain, and $M$ is the identified number of modes. The evaluation
penalizes frequency, decay, gain, and mode-count errors.

Existing estimators broadly follow either direct signal analysis or
analysis-by-synthesis. Sinusoidal analysis estimates resonant components from
the signal~\cite{mcaulay1986sinusoidal}, while subspace and matrix-pencil methods
estimate damped exponentials when the model order and signal conditions are
suitable~\cite{roy1989esprit,hua1990matrix}. Differentiable digital signal
processing (DDSP) instead allows a structured signal model to be fitted by
gradient descent~\cite{engel2020ddsp,diaz2023rigid}, but a large variable-size
modal set still requires an initial structure. The official reference method
uses spectral peak picking followed by local decay and gain estimates. It is
inexpensive, but sparse peak detection can miss weak or overlapping modes.
The specific challenge is therefore not only to calibrate individual modes,
but also to construct a modal set with a plausible cardinality.

This paper asks whether band-wise mode-count estimation can provide the
structural prior needed before local modal calibration. We propose a count-first
system that predicts mode counts in broad frequency bands, places a dense
modal grid, and applies bounded differentiable refinement to decay and gain.
This decomposes variable-size estimation into learned count prediction,
deterministic initialization, and constrained fitting. On two simulator-matched
validation sets, the system substantially lowers local challenge-style error
relative to the official default baseline. Most of the gain comes from reducing
mode-count mismatch, while differentiable refinement adds a smaller improvement.

\section{Method}

\begin{figure*}[t!]
\centering
\includegraphics[width=0.98\textwidth]{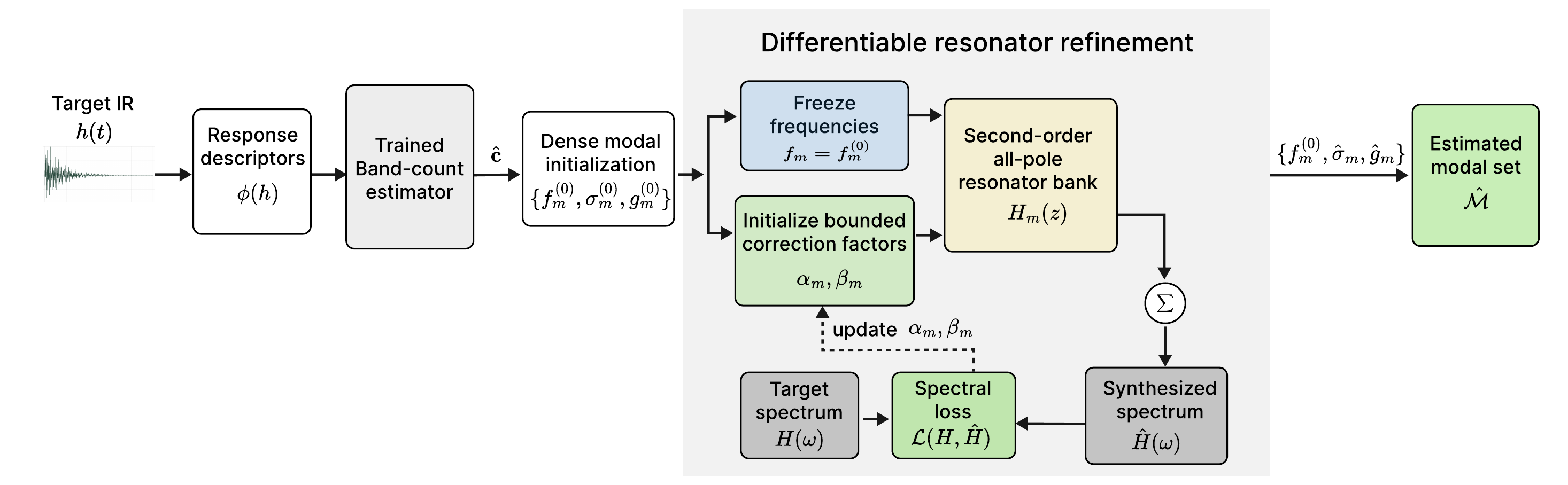}
\caption{Overview of Task B inference. Fixed response descriptors feed the
trained band-count estimator, whose predictions define a dense modal
initialization. A spectral objective through the differentiable resonator bank
then updates bounded decay and gain correction factors while keeping $f_m$
fixed. Offline training of the count estimator is described in the text.}
\label{fig:pipeline}
\end{figure*}

We use a hybrid computational inverse-estimation design because Task B requires
both an unknown number of modes and continuous parameters for every mode.
Supervised learning estimates band-wise mode counts, deterministic initialization
places the frequencies and estimates decay and gain from the response, and
constrained per-response optimization refines decay and gain. Offline, each
simulated response
$h_n^{\mathrm{syn}}$ is paired with its ground-truth modal list
$\mathcal{M}_n^{\mathrm{gt}}$. A band-histogram operator $B$ converts the list
to four count labels,
\begin{equation}
(h_n^{\mathrm{syn}},\mathcal{M}_n^{\mathrm{gt}})
\mapsto
(\phi(h_n^{\mathrm{syn}}),\mathbf c_n),
\quad \mathbf c_n=B(\mathcal{M}_n^{\mathrm{gt}}),
\end{equation}
where $\phi$ is a deterministic 372-dimensional descriptor extractor with no
learned parameters and is identical at training and inference. Robust-scaling
statistics and four ExtraTrees regressors are fitted from these pairs. We denote
the complete fitted predictor by $F_{\mathrm{ET}}:\phi(h)\mapsto\mathbf c$.
At inference, $F_{\mathrm{ET}}$ and the deterministic initializer $I$ produce
\begin{equation}
\hat{\mathbf c}=F_{\mathrm{ET}}(\phi(h)),\qquad
I(h,\hat{\mathbf c})=
\{f_m^{(0)},\sigma_m^{(0)},g_m^{(0)}\}_{m=1}^{\hat M},
\end{equation}
where $\hat M$ is the sum of the bounded, rounded band counts. The final stage
compares the target complex spectrum $H(\omega)$ with the resonator-bank
spectrum $\hat H(\omega)$ and solves the constrained problem
\begin{align}
(\hat{\boldsymbol\alpha},\hat{\boldsymbol\beta})
=\arg\min_{\boldsymbol\alpha,\boldsymbol\beta}
&\ \mathcal{L}\!\left(H,
\hat H(\mathbf f^{(0)},
\boldsymbol\sigma^{(0)}\!\odot\!\boldsymbol\alpha,
\mathbf g^{(0)}\!\odot\!\boldsymbol\beta)\right) \nonumber\\
\text{subject to}\quad
&\alpha_m,\beta_m\in[1/8,8],\qquad f_m=f_m^{(0)},
\end{align}
where bold symbols collect the per-mode quantities and $\odot$ denotes
element-wise multiplication. The method returns
\begin{equation}
\hat{\mathcal M}=
\{(f_m^{(0)},\sigma_m^{(0)}\hat\alpha_m,
g_m^{(0)}\hat\beta_m)\}_{m=1}^{\hat M}.
\end{equation}
Only the target response is required at inference. The simulator and reference
modal list are not used. Figure~\ref{fig:pipeline} summarizes the inference
pipeline.

\subsection{Band-wise mode counts}

The frequency range is split into four bands: 20--200 Hz, 200--1000 Hz,
1--4 kHz, and 4--10 kHz. A four-output estimator predicts the number of modes
in each band from 372 response descriptors. They summarize unnormalized
amplitude, early/late energy, spectral shape, peak spacing, group delay, and
band-limited decay, and are fixed rather than learned. After robust feature
scaling, each output is predicted by an independent ExtraTrees
regressor~\cite{geurts2006extremely} with 500 trees and a minimum leaf size of 2.

Predicting mode counts per band lets the representation adapt its density to
each target response rather than using a single fixed number of modes over the
entire spectrum. This stage is the only supervised learning component in the
Task B system.

\subsection{Dense modal initialization}

Given the predicted count in each band, modal frequencies are placed on a
dense linear grid. Predicted counts are rounded, with at least eight modes per
band and a 15,000-mode total cap. This is a count-conditioned approximation,
not an analytical plate-mode enumeration: it uses neither the plate
eigenfrequency equation nor explicit $(m,n)$ mode indices. Let $q$, in
$\mathrm{dB\,s^{-1}}$, be the slope of a linear fit to the energy decay curve
between $-5$ and $-35$ dB. With at least eight samples in this range, we set
$\sigma_{\mathrm{base}}=\max(0.1,-q\log(10)/20)$. Otherwise, we use
$\sigma_{\mathrm{base}}=2$. All decay rates are expressed in $\mathrm{s}^{-1}$.
A mode in band $b$ is initialized as
\begin{equation}
\sigma_m^{(0)}=\sigma_{\mathrm{base}}s_b
\left(1+\frac{0.5f_m^{(0)}}{c_b}\right),
\end{equation}
where $\mathbf s=[20,20,20,40]$ contains dimensionless band scales and $c_b$
is the band center in Hz. The larger final-band scale and the factor 0.5 impose
stronger high-frequency decay and a mild within-band increase, respectively.
Initial gains are heuristically computed from the
linearly interpolated complex DFT $H$ at each grid frequency,
\begin{equation}
g_m^{(0)}=-2\,\mathrm{Im}\{H(2\pi f_m^{(0)})\}\,\sigma_m^{(0)}T
\sin(2\pi f_m^{(0)}T).
\end{equation}
Here, $T$ is the sampling period.

\subsection{Differentiable resonator refinement}

The initializer is refined with a differentiable bank of second-order all-pole
resonators. This follows the broader differentiable digital signal processing
idea that a structured signal model can be optimized by gradient
descent~\cite{engel2020ddsp}, while using a resonator form standard in modal sound
synthesis and physical audio modeling~\cite{smith2010physical,bilbao2009numerical}.
At angular frequency $\omega$, the modeled response is
\begin{equation}
\hat{H}(\omega)=
\sum_m
\frac{g_m z}
{1-2r_m\cos(\omega_m T)z+r_m^2z^2},
\quad
r_m=e^{-\sigma_m T},
\end{equation}
where $z=e^{j\omega T}$ and $\omega_m=2\pi f_m$.

In the final system, the mode count and $f_m$ values are fixed by the
dense initializer. Per-mode $\sigma_m$ and $g_m$ are updated through bounded
multiplicative correction factors, so this stage acts as local calibration
rather than full modal re-identification. For unconstrained optimizer variables
$a_m$ and $b_m$, we use
\begin{equation}
\alpha_m=e^{\log(8)\tanh(a_m)},\qquad
\beta_m=e^{\log(8)\tanh(b_m)},
\end{equation}
with $a_m=b_m=0$ initially. Thus both corrections start at one, remain within
the eightfold trust region, and preserve the initial gain sign. The objective
combines log-magnitude matching, a small phase-consistency term, and
regularization:
\begin{align}
\mathcal{L}
=&\ \underset{\omega}{\operatorname{mean}}
\left|\log(|\hat H|+\epsilon)-\log(|H|+\epsilon)\right| \nonumber\\
&+\lambda_{\mathrm{ph}}\underset{\omega}{\operatorname{mean}}
\left|u(\hat H)-u(H)\right| \nonumber\\
&+\lambda_\sigma R_\sigma+\lambda_g R_g ,
\end{align}
where $\epsilon=10^{-12}$, $u(H)=H/(|H|+\epsilon)$ is the unit complex spectrum,
$R_\sigma=\operatorname{mean}_m[(\log\alpha_m)^2]$, and
$R_g=\operatorname{mean}_m[(\log\beta_m)^2]$. The final setting uses
$\lambda_{\mathrm{ph}}=\lambda_\sigma=\lambda_g=0.02$. At each step, the resonator bank
produces $\hat H$ on 1024 frequency samples from 20 Hz to 10 kHz. The loss is
backpropagated through the bank to the decay and gain correction variables.
frequency is not an optimization variable. Adam runs for 80 steps with learning
rate 0.02 and gradient-norm clipping at 1, and the lowest-loss iterate is
retained. The count estimator is implemented with scikit-learn, and the
resonator refinement uses PyTorch automatic differentiation.

\section{Experiments}

Because the official modal lists are hidden, all development uses data generated
by the public simulator. The count estimator uses 600 synthetic 5 s responses
with a fixed 450/150 training/validation split. Modal-set accuracy is evaluated
on two separately generated, disjoint 5 s sets: \emph{Validation 1} contains
8 responses and \emph{Validation 2} contains 12. Both are used to compare
refinement bounds.

For modal-set experiments we report a local challenge-style score
\begin{equation}
\mathrm{RE}=\mathrm{RE0}+|M-\hat M|/M, \qquad
\mathrm{RE0}=(\mathrm{RE}_f+\mathrm{RE}_\sigma+\mathrm{RE}_g)/3,
\end{equation}
where $M$ and $\hat M$ are the reference and estimated mode counts. The first
term averages the frequency, decay, and gain errors, while the second penalizes
mode-count mismatch. Lower is better. Component-wise relative errors are clipped
at one, and unmatched reference modes also contribute an error of one. To handle
dense lists, modes are paired by greedy monotone frequency matching with a
0.5-octave threshold. This approximation is used only for local model selection.
It is not the official hidden-label score.

The official default peak picker is the reference baseline. It uses 6 dB
prominence, 2 Hz minimum spacing, and a 2 Hz prominence half-window over the same
20 Hz--10 kHz range. All modal-set variants use the same local matcher and score.
For stage ablation, a total-count dense initializer predicts one mode count and
allocates it to the four bands using fixed training-set fractions. We evaluate
this initializer before and after fourfold decay/gain refinement, then replace
it with the proposed direct band-count estimator and vary the refinement bound.

Runtime over the 16 released responses was measured on a MacBook Pro with an
Apple M1 Pro chip (10-core CPU, 16 GB unified memory) running macOS 14.5, without
GPU acceleration. Timing excludes synthetic-data generation and count-model
training.

\section{Results}

Table~\ref{tab:taskb-val} compares the complete system with the official
default peak-picking baseline and separates the main system stages. Fourfold
refinement improves the total-count dense initializer on both validation sets,
and direct band-count prediction lowers RE further. Increasing the correction
bound from fourfold to eightfold again improves both sets, whereas a sixteenfold
bound does not transfer consistently. The eightfold setting is therefore
retained. It reduces local RE by about 66\% relative to the reference baseline
on both validation sets.

\begin{table}[ht]
\caption{Local Task B relative error and stage ablation. Lower is better. Bold
marks the column minimum.}
\centering
\setlength{\tabcolsep}{5pt}
\begin{tabular}{lcc}
\toprule
Method & Validation 1 & Validation 2 \\
\midrule
Official peak picking (default) & 1.9699 & 1.9686 \\
Total-count dense initialization & 0.7364 & 0.7398 \\
$+$ refinement (4$\times$) & 0.6859 & 0.6935 \\
Band-count $+$ refinement (4$\times$) & 0.6738 & 0.6796 \\
Band-count $+$ refinement (8$\times$) & 0.6651 & \textbf{0.6750} \\
Band-count $+$ refinement (16$\times$) & \textbf{0.6627} & 0.6881 \\
\bottomrule
\end{tabular}
\label{tab:taskb-val}
\end{table}

A major discrepancy in the reference baseline is mode count. It
identifies an average of 69.0 and 67.3 modes on Validation 1 and 2,
respectively, whereas the corresponding reference lists contain averages of
5240.1 and 6071.9 modes. The count-first system reduces the normalized count
mismatch to about 5--6\%.

On the separate 150-response count-validation split, ExtraTrees reduces both
mean per-band and total-count MAE relative to Random Forest
(Table~\ref{tab:count-val}), so it is retained in the complete system.

\begin{table}[ht]
\caption{Mode-count model selection on the 150-response count-validation split.}
\centering
\begin{tabular}{lcc}
\toprule
Model & Band MAE & Total MAE \\
\midrule
Random forest & 75.11 & 300.12 \\
ExtraTrees & \textbf{67.64} & \textbf{270.41} \\
\bottomrule
\end{tabular}
\label{tab:count-val}
\end{table}

Table~\ref{tab:component-val} decomposes the selected system's remaining error.
Normalized count mismatch is about 0.05, frequency error about 0.31, and decay
and gain errors range from 0.69 to 0.85.

\begin{table}[ht]
\caption{Component errors for the selected eightfold setting.}
\centering
\setlength{\tabcolsep}{3.5pt}
\begin{tabular}{lccccc}
\toprule
Split & RE0 & $\mathrm{RE}_f$ & $\mathrm{RE}_\sigma$ & $\mathrm{RE}_g$ & $|\Delta M|/M$ \\
\midrule
Validation 1 & 0.6087 & 0.3176 & 0.6912 & 0.8172 & 0.0564 \\
Validation 2 & 0.6215 & 0.3081 & 0.7015 & 0.8548 & 0.0536 \\
\bottomrule
\end{tabular}
\label{tab:component-val}
\end{table}

For the final predictions on the 16 released responses, the mean refinement
time was \SI{6.51}{s} per response and end-to-end inference took
\SI{105.64}{s}. The output contains approximately 83,800 estimated modes in
total, with individual counts ranging from roughly 1,100 to 14,200.

\section{Discussion}

The comparison indicates that avoiding severe undercounting is central to the
improvement over the official reference method. Peak picking returns only
locally prominent resonances, whereas the plate simulator produces thousands of
modes, including weak and overlapping components. Predicting counts in broad
bands therefore provides a useful density prior before per-mode fitting.
ExtraTrees improves the held-out count errors over Random Forest, although the
remaining mismatch shows that count estimation is not solved completely.

Changing the decay/gain correction bound has a smaller effect than replacing
sparse peak picking with count-first estimation. Increasing the bound from
fourfold to eightfold improves both validation sets, while a sixteenfold range
does not transfer consistently. This suggests that bounded calibration can
correct part of the approximate, signal-informed initialization, but excess
freedom can worsen the fit. Frequency error is unchanged because the grid is
fixed, and decay and gain remain the dominant component errors.

Together, the findings support a hybrid division of roles: a learned estimator
maps fixed descriptors to modal density, while a structured resonator model
calibrates continuous parameters. This complements subspace methods that require
a suitable model order~\cite{roy1989esprit,hua1990matrix} and DDSP methods that
optimize structured synthesizers~\cite{engel2020ddsp,diaz2023rigid}. Here, the
learned stage supplies the variable-size structure needed before differentiable
fitting.

The evidence is limited to small validation sets generated by the same
simulator family, and both sets were used when selecting the correction bound.
The greedy monotone matcher is a memory-safe approximation rather than the
official hidden-label evaluator, so the reported scores do not establish the
challenge ranking. The official baseline comparison also does not cover tuned
peak picking, subspace or matrix-pencil estimation, or unconstrained
differentiable fitting. Finally, a misaligned grid frequency cannot be corrected
by the present refinement. Future work should combine learned count priors with
subspace-derived candidates or bounded frequency updates, improve decay/gain
initialization, and estimate count uncertainty. Modal-list errors should also be
compared with resynthesized response and perceptual measures, since these
criteria need not improve proportionally.

\section{Conclusion}

This paper examined dense modal estimation from plate-reverb impulse responses.
On local synthetic validation, predicting band-wise mode counts before fitting
substantially reduces error relative to the official default peak-picking
baseline, while bounded decay/gain refinement gives a smaller additional improvement.
The results show that modal density is an important part of estimating large
sets of weak and overlapping modes. Future work should allow limited frequency
updates, improve decay and gain estimation, and compare with stronger classical
modal-estimation methods and the official hidden-label evaluation.

\bibliographystyle{IEEEtranDAFx}
\bibliography{references}

\end{document}